\documentclass[prb,reprint,showpacs,amsmath,amssymb,superscriptaddress]{revtex4-1}
\usepackage{graphicx}
\usepackage{xcolor}
\usepackage[colorlinks,bookmarks=false,citecolor=blue,linkcolor=red,urlcolor=blue]{hyperref}
\usepackage{multirow}
\emergencystretch=\maxdimen
\begin{document}
\title{One-dimensional Hubbard-Holstein model with finite range
  electron-phonon coupling}

\author{F. H\'ebert}
\email[Corresponding author: ]{frederic.hebert@inphyni.cnrs.fr}
\affiliation{Universit\'e C\^ote d'Azur, CNRS, INPHYNI, France}
\author{Bo Xiao}
\affiliation{Department of Physics, University of California, Davis,
  California 95616, USA} 
\author{V.$\,$G.  Rousseau}
\affiliation{Physics Department, Loyola University New Orleans, 6363
  Saint Charles Ave., LA 70118, USA} 
\author{R$\,$T. Scalettar}
\affiliation{Department of Physics, University of California, Davis,
  California 95616, USA} 
\author{G.$\,$G. Batrouni}
\affiliation{Universit\'e C\^ote d'Azur, CNRS, INPHYNI, France}
\affiliation{MajuLab, CNRS-UNS-NUS-NTU International Joint Research
  Unit UMI 3654, Singapore}
\affiliation{Department of Physics, National University of Singapore, 2 Science Drive 3, 117542 Singapore}
\affiliation{Centre for Quantum Technologies, National University of
  Singapore; 2 Science Drive 3 Singapore 117542} 
\affiliation{Beijing Computational Science Research Center, Beijing
  100193, China}

\begin{abstract}
The Hubbard-Holstein model describes fermions on a discrete lattice,
with on-site repulsion between fermions and a coupling to phonons that
are localized on sites. Generally, at half-filling, increasing the coupling $g$ to the phonons drives
the system towards a Peierls charge density wave state whereas increasing the
electron-electron interaction $U$ drives the fermions into a Mott
antiferromagnet. At low $g$ and $U$, or when doped, the system is
metallic.  In one-dimension, using quantum Monte Carlo simulations, we
study the case where fermions have a long range coupling to phonons,
with characteristic range $\xi$,
interpolating between the Holstein and Fr\"ohlich limits. Without
electron-electron interaction, the fermions adopt a Peierls state when the
coupling to the phonons is strong enough. This state is destabilized
by a small coupling range $\xi$, and leads to a collapse of the fermions,
and, consequently, phase separation.  Increasing interaction $U$ will
drive any of these three phases (metallic, Peierls, phase separation)
into a Mott insulator phase. The phase separation region is once
again present in the $U \ne 0$ case, even for small values of the
coupling range.
\end{abstract}


\maketitle

\section{Introduction}

Coupling between electrons and phonons is ubiquitous in solid state
physics, resulting in many important phenomena such as polarons
\cite{Frohlich54}, effective Cooper pairing between electrons
\cite{Cooper56}, or to density modulations such as the Peierls
instability \cite{Peierls79}.
The Holstein model \cite{Holstein59} is a simple model describing such
coupling.  It is especially amenable to numerical treatment since it
describes phonons as localized particles that interact locally with
free fermions on a lattice.  At half-filling, the
Holstein model exhibits a transition between an homogeneous metallic
phase and a gapped charge density wave (CDW) Peierls insulating
phase \cite{Scalettar89,Marsiglio90,Freericks93,McKenzie96,Bursill98}. An effective attraction between fermions, mediated by phonons,
triggers this instability for large enough electron-phonon
coupling\cite{Hirsch82,Fradkin83,Jeckelmann99,Hardikar07,
  Greitemann15}.  In this work, we will concentrate on the
one-dimensional version of the model.

Many effects are not taken into account in the original Holstein model
that can alter the physics of fermion-phonon systems. Non-local
coupling between fermions and phonons is expected in some materials
and leads to the interpolation between Holstein's local description
and Fr\"ohlich's description where electrons and phonons interact at
long distances\cite{Frohlich54}.  This problem has been studied in the
context of polaron formation\cite{Devreese09,Alexandrov99,Fehske00,Chandler14},
high temperature superconductivity \cite{Hardy09}, and
recently\cite{Hohenadler12} for its impact on the physics of Peierls
instability.  It was shown\cite{Hohenadler12} that increasing the
coupling range leads to a collapse of the fermions causing them to
clump together in one part of the system, {\it i.e.} phase separation.

Direct interactions between fermions are not included in the Holstein
model, but a variant, dubbed the Hubbard-Holstein model
\cite{Beni74,Takada03}, includes local interactions between
fermions. At half filling, onsite interactions drive the system into
an antiferromagnetic (AF) Mott insulator but there is competition
between the Peierls and Mott phases, than can lead to the appearance
of an intermediate metallic phase
\cite{Takada95, Takada96,Fehske02,Tezuka05,Tezuka07,Hardikar07,Fehske08,Hohenadler13,Nocera14}.

The goal of this article is to study both the effects of the long range
e-p (electron-phonon) coupling and those of direct e-e
(electron-electron) repulsion in a one dimensional system.  This leads
to a rich phase diagram at half-filling where four competing phases
come into play: metallic, Peierls, Mott phases and phase separation.
Other modifications of the Hubbard-Holstein model have been envisioned
such as an anharmonicity of the phonons \cite{Lavanya17}, or the
effect of different band structures on the pairing of the fermions
\cite{Tezuka05}.

The paper is organised as follow. First, we introduce the model and
the quantum Monte Carlo (QMC) methods. Then
we study the system with long range e-p coupling but without e-e
interactions to validate our approach and compare with other
work \cite{Hohenadler12}. Finally, the main results concerning the
system with both \mbox{e-e} interactions and long range e-p coupling will be
presented and compared with results obtained in the on-site coupling
limit \cite{Hardikar07, Fehske08, Tezuka07}.

\section{Hamiltonian and Methods}
\label{sec_2}

\subsection{Model}
\label{sec2_subA}

We consider the following model
\begin{eqnarray}
H   =&&   - t \sum_{r,\sigma}
\left(  c^\dagger_{r,\sigma} c^{\phantom\dagger}_{r+1,\sigma} + {\rm H.c.}
\right ) + U  \sum_r n_{r,\uparrow}n_{r,\downarrow}\nonumber\\
 && + \omega \sum_r n_{r,\phi}
   + \sum_{r,R} G(R) \sqrt{2}\, X_r n_{r+R}
\label{Hamiltonian}
\end{eqnarray}
The fermionic operators $c^\dagger_{r,\sigma}$ and
$c^{\phantom\dagger}_{r,\sigma}$ respectively create and destroy a
fermion with spin $\sigma = \uparrow,\downarrow$ on site $r$ of a one
dimensional periodic lattice containing $L$ sites. Similarly,
$a^\dagger_r$ and $a_r$ are phonon creation and annihilation operators
on site $r$. The operators $n_{r,\sigma} = c^\dagger_{r,\sigma}
c^{\phantom\dagger}_{r,\sigma}$, $n_r = n_{r,\uparrow} +
n_{r,\downarrow}$ and $n_{r,\phi} = a^\dagger_{r}a_{r}$ represent, on
site $r$, the number of fermions of spin $\sigma$, the total number of
fermions and the number of phonons, respectively. The corresponding
densities will be noted $n_\uparrow$, $n_\downarrow$, $n = n_\uparrow+n_\downarrow$ and $n_\phi$
(for example, $n_\phi = \sum_r \langle n_{r,\phi}\rangle / L$).

The first (second) term of Eq.~(\ref{Hamiltonian}) describes the fermionic
kinetic (potential) energy; together they give the conventional
fermionic Hubbard model. The hopping parameter sets the energy scale,
$t=1$. The additional terms are the diagonal energy of phonons with
frequency $\omega$, and the coupling between the displacement of the
lattice at position $r$, $X_r = (a^\dagger_{r}+ a_{r})/\sqrt{2}$, and the density
of fermions at site $r+R$, which describes long range electron-phonon
coupling. The coupling $G(R)$ is characterised by its overall
strength $g$ and its range $\xi$ and is given by
\begin{equation}
G(R) = g\frac{\exp(-|R|/\xi)}{\left(1+R^2\right)^{3/2}}.
\end{equation}
Due to periodic boundary conditions, $R$ is defined as the minimum of
$R$ and $L-R$.

\subsection{Methods}
\label{sec2_subB}

We study the Hamiltonian Eq.~(\ref{Hamiltonian}), in the cases where
the electrons are interacting with each others ($U\ne 0$) and where they are not ($U=0$), focusing on
one value of $\omega =t/2$. To this end we use the directed stochastic
Green function algorithm\cite{Rousseau08} (SGF) which allowed us to
simulate systems with size up to $L=42$. The inverse temperature
$\beta$ was typically chosen proportional to the size of the lattice
$\beta t = L$, which we found to be large enough to ensure convergence to ground state
properties. The algorithm uses the mapping of fermionic degrees of
freedom onto hardcore bosons using the Jordan-Wigner transformation
\cite{Jordan28}. The convergence to equilibrium is sometimes quite
difficult, especially when the system undergoes phase separation. To
circumvent this problem, we performed simulations with different initial
conditions and accept the results corresponding to the lowest free
energy. In the most difficult cases, we could obtain reliable results
on sizes only up to $L=18$, which does not allow a complete finite
size scaling analysis of the phase transitions.

To verify the SGF results, the Hamiltonian was also studied for $U=0$
using a new algorithm based on a Langevin simulation technique
initially used for lattice field theories
\cite{Batrouni85,Batrouni19}. The algorithms are presented in more detail 
in appendix \ref{appendix}.

We will use static quantities and correlation functions to analyse the
system.  The fermion densities, $n_\sigma$, are fixed in the
canonical SGF algorithm while the density of phonons $n_\phi$
fluctuates due to the $X_r$ term in the Hamiltonian. We concentrate on
the half-filled case where $n_\uparrow = n_\downarrow = 1/2$.

The one particle Green functions $G_\sigma(R)$, and $G_\phi(R)$ probe
the phase coherence of the different kinds of particles. They are
defined as
\begin{eqnarray} 
G_\sigma(R) &=& \langle c^\dagger_{r+R,\sigma} c^{\phantom{\dagger}}_{r,\sigma} + {\rm
  H.c.} \rangle/2, \nonumber \\ G_\phi(R) &=& \langle a^\dagger_{r+R}
a^{\phantom{\dagger}}_{r} + {\rm H.c.} \rangle/2.
\end{eqnarray} 
For the fermions, we have an indirect access to the phase stiffness
through the Jordan-Wigner mapping to hardcore bosons: as for fermions
Green function becomes long ranged or quasi-long ranged,
it also does so for hardcore bosons and the phase stiffness (superfluid
density) of the bosons becomes non zero. This stiffness is calculated
by the fluctuations of the winding number of the bosons:
$\rho_{s,\sigma} = \langle W_\sigma^2 \rangle/2\beta t$.

To identify the Peierls phase, we use density-density correlations
$D_{\alpha\beta} (R) = \langle n_{r,\alpha} n_{r+R,\beta} \rangle -
n_\alpha n_\beta$, where $\alpha$ and $\beta$ correspond to particles
species (electrons or phonons). The corresponding structure factors,
$F_{\alpha\beta}(k)$, which are the Fourier transforms of the
density-density correlations functions, are given by
\begin{equation}
F_{\alpha\beta}(k) = \sum_R \langle n_{r,\alpha} n_{r,\beta}\rangle
\exp(ikR)/L.
\end{equation}
$k$ varies in the interval $\lbrack-\pi,\pi \rbrack$ with step size
$\epsilon = 2 \pi/L$.  The Peierls phase, with alternating empty and
filled sites shows pronounced peaks of the structure factors at
$k=\pi$.

Finally, in the case where $U$ is different from zero, we expect some antiferromagnetic
correlations to appear, which we will identify by using $S_{zz}(\pi)$, the
Fourier transform of the spin-spin correlations along the $z$ axis,
\begin{equation}
S_{zz}(k) = \sum_R \langle S_{r,z} S_{r+R,z} \rangle \exp(ikR)/L.
\end{equation}
where $S_{r,z} = (n_{r,\uparrow} - n_{r,\downarrow})/2$.  
The spin correlations in the $xy$ plane should be the same
as the system has a $SU(2)$ symmetry and, with this continuous symmetry,
we expect only quasi long range ground state order for the spin correlations.

In terms of hardcore bosons, the AF phase transforms into a Mott phase with two
species of bosons. The AF correlations in the $xy$ plane
correspond to counter-superfluid correlations\cite{Kuklov03,Pollet06}.
In such a state, we have boson-hole quasi particles, corresponding to boson
exchanges, that show quasi long range phase order and give a non zero stiffness
$\rho_{s,\sigma}$, despite the fact that individual particles are exponentially
localized in the Mott phase.
In the Mott AF phase, individual Green functions $G_\sigma$
decay exponentially but $\rho_{s,\sigma}$ is still expected to be non zero,
due to quasi long range spin correlations in the $xy$ plane.

\section{Case with zero electron-electron interaction}

The $U=0$ case has already been studied in
Ref.~[\onlinecite{Hohenadler12}] where it was shown that, at
half-filling $n_\uparrow = n_\downarrow =1/2$, the system exhibits
three phases: A metallic phase with quasi long range order for density
and phase correlations, a Peierls phase with a charge order with sites
alternately occupied by particles or almost empty and where movement
of fermions is suppressed, and finally, when $\xi$ is large enough,
phase separation between regions that are almost completely
filled with fermions ($\langle n_{r,\uparrow}\rangle = \langle
n_{r,\downarrow} \rangle = 1$) and regions that are empty.

\begin{figure}[!h]
\includegraphics[width=1 \columnwidth]{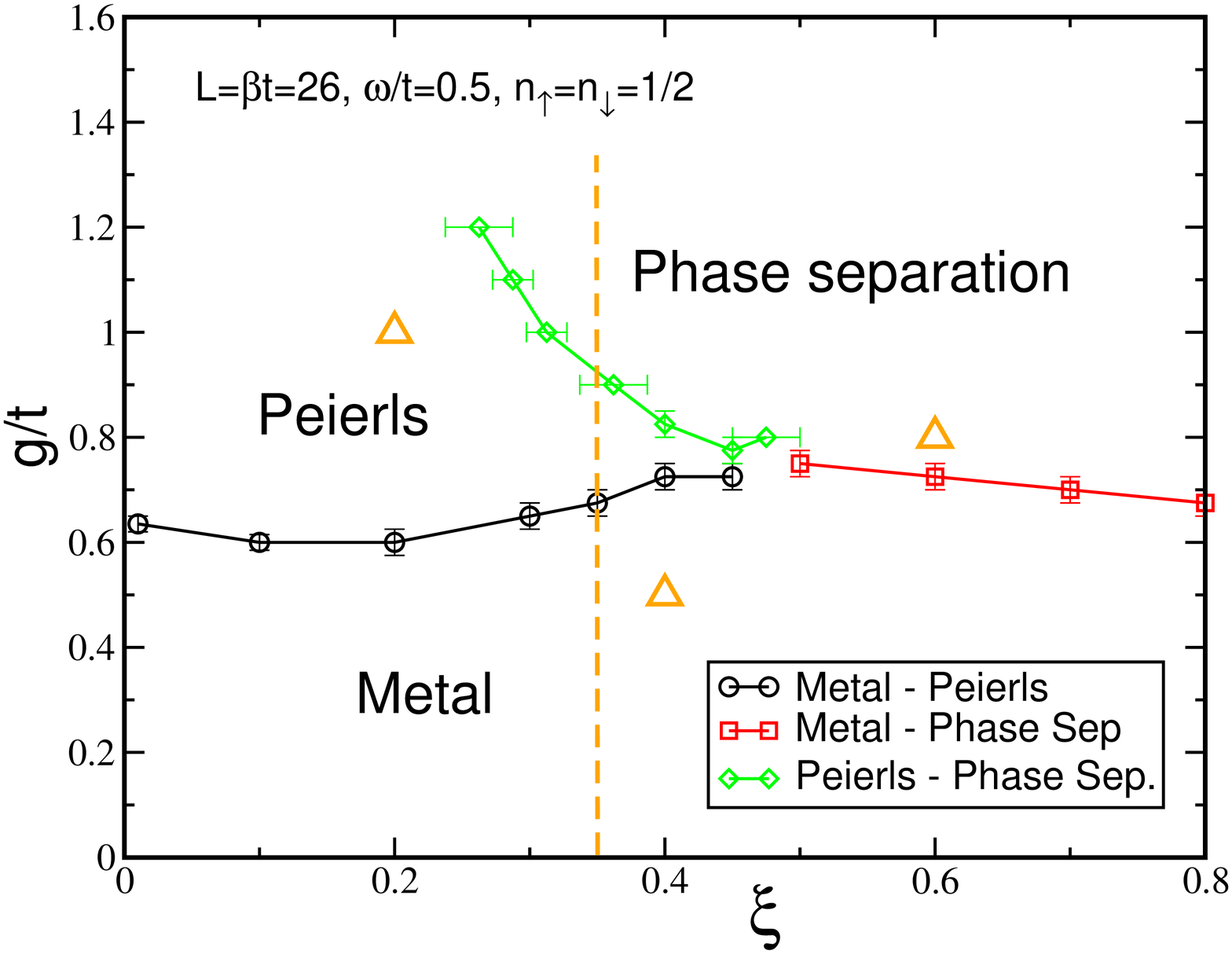}
\caption{(Color online) Phase diagram for the $U=0$
  system as a function of the coupling strength, $g$, and range,
  $\xi$, of the fermion-phonon coupling.  The triangles correspond
  to three cases which are studied in detail in the text. The dashed
line corresponds to the cut shown in Fig.~\ref{fig:cutxifixe}.
\label{fig:phase_diagram_U0}}
\end{figure}

As shown in Fig.~\ref{fig:phase_diagram_U0}, we found the same three
phases. However, our phase diagram is quite different from that of
Ref.~[\onlinecite{Hohenadler12}].  In particular, we observe phase
separation for $\xi$ below 0.3 whereas, in
[\onlinecite{Hohenadler12}], it was only observed for $\xi \ge 2$. To
explain how we constructed the phase diagram,
Fig.~\ref{fig:phase_diagram_U0}, we will first present the properties
of the three different phases, using as examples the three points
(triangles) represented in Fig.~\ref{fig:phase_diagram_U0}.

\begin{figure}[!h]
\includegraphics[width=1 \columnwidth]{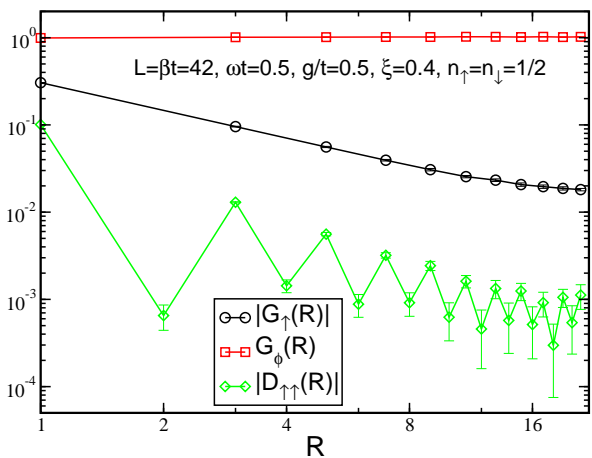}
\caption{(Color online) The one body Green functions for up fermions,
  $G_\uparrow (R)$, phonons, $G_\phi (R)$, and the fermion
  density-density correlations, $D_{\uparrow\uparrow}(R)$, as functions of distance in
  the metallic phase. Logarithmic scales are used on both axes. The phonon Green function, $G_\phi$, shows long
  range order while for fermions, $G_\uparrow$ and
  $D_{\uparrow\uparrow}$ show quasi-long range order.
\label{fig:metal}}
\end{figure}

In Fig.~\ref{fig:metal}, we show the Green functions for the up
fermions, $G_\uparrow$, and phonons, $G_\phi$, as well as fermion
density-density correlations, $D_{\uparrow\uparrow}$, in the metallic
phase for $\xi=0.4$ and $g/t=0.5$.  As expected in a one-dimensional
system \cite{Giamarchi03}, $G_\uparrow$ and $D_{\uparrow\uparrow}$
show an algebraic decay with distance $R$, typical of quasi long range
order. Here the dominant effect is the one particle
motion, as $G_\uparrow$ decays is slower than for
$D_{\uparrow\uparrow}$.  On the contrary, $G_\phi$ shows that the
phonons adopt a long range ordered phase. This is expected as the
coupling to the fermion density acts as an external field for the
phonon displacement, $X_r$, and thus provides an explicit symmetry
breaking.  A simple coherent state approximation shows that, for a
homogeneous density of fermions $n$, $\phi = \langle a \rangle =
-n\sum_R G(R)/\omega$ and $n_\phi = G_\phi = \phi^2$. This simple
ansatz yields $G_\phi = n_\phi = 1.12$ for the case considered here,
whereas our simulation gives a slightly lower value $G_\phi(L/2) =
1.016 \pm 0.007$. As the number of phonons increases, the coherent
state approach describes the system more accurately. For example, for
$\xi=g/t=0.6$, it predicts $n_\phi = G_\phi = 1.87$ while the
numerical value is $G_\phi(L/2) = 1.91\pm 0.02$.

\begin{figure}[!h]
\includegraphics[width=1 \columnwidth]{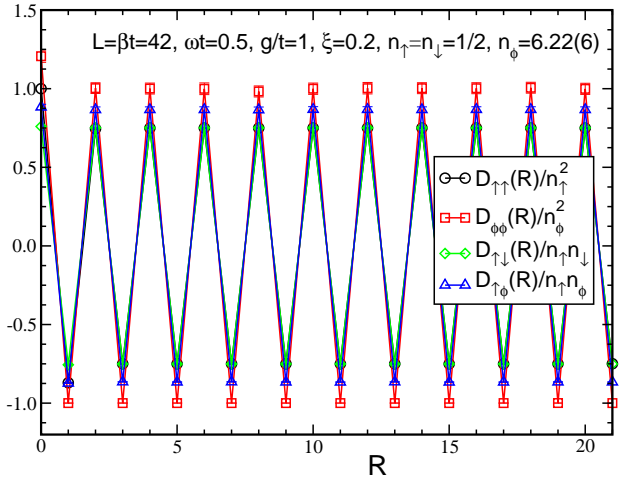}
\caption{(Color online) Behavior of different density-density
  correlation functions with distance in the Peierls phase. The
  Peierls phase shows an alternation of sites that are occupied by
  fermions and phonons with sites that are almost empty.  The
  functions have been rescaled by respective densities for better
  visibility.
\label{fig:peierlsdens}}
\end{figure}

Turning now to the Peierls phase, we observe that, when $g$ is large
enough ($\xi=0.2, g/t=1.0$) the homogeneous metallic phase is
destabilised and changes into a Peierls state with a modulation of
densities (a charge density wave, CDW).  All the density-density
correlation functions exhibit the same characteristic oscillations
with wave vector $k=\pi$ (Fig.~\ref{fig:peierlsdens}).  Following the
previous ansatz, the coupling energy between the fermions and the
phonons on a site is roughly proportional to $-n_r^2 g^2 /\omega$.  The
electron-phonon coupling energy will then be multiplied by
approximately two when the system undergoes a transition from a
homogeneous phase where $n_r \simeq 1$ on each site to a state where
$n_r \simeq 2$ every other site. 
  This happens for large enough $g$ as the transition increases
  the hopping energy: delocalized particles occupy 
long wavelength states with negative energies, while
localized particles have a hopping energy which is approximately zero.
The decrease of the coupling energy should then compensate for this hopping energy increase.

The alternation of occupied and empty sites is similar to what is observed in the
  attractive Hubbard model, with the major difference that the
  attractive effect between fermions is mediated by the phonon field.
  The CDW structure is stabilized as it offers the largest amount of
  virtual hopping possibilities for the fermions.

\begin{figure}[!h]
\includegraphics[width=1 \columnwidth]{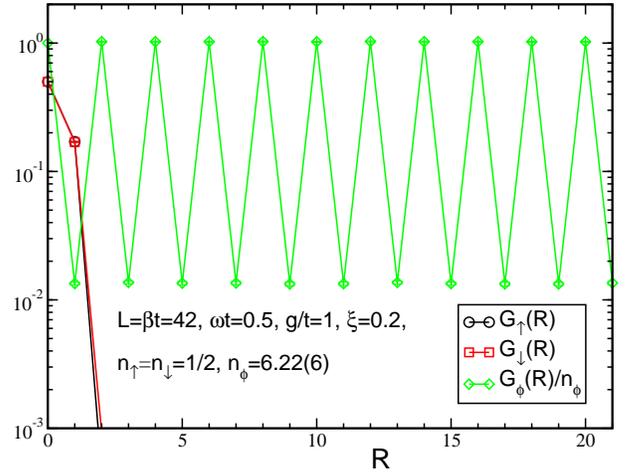}
\caption{(Color online) Behavior of Green functions in the Peierls
  phase. The fermion Green functions, $G_\uparrow(R)$ and
  $G_\downarrow(R)$, decay rapidly to zero while the phonon Green
  function, $G_\phi(R)$, goes to a plateau when $R$ becomes
  large. This plateau is modulated by the CDW density oscillations.
\label{fig:peierlsgreen}}
\end{figure}

The localization of the fermions is immediately visible in the
behaviour of $G_\uparrow(R)$ and $G_\downarrow(R)$ which decay
exponentially. However, $G_\phi(R)$ still shows a plateau at long
distances, which shows the condensation of phonons
(Fig. \ref{fig:peierlsgreen}), albeit modulated by the density wave.

\begin{figure}[!h]
\includegraphics[width=1 \columnwidth]{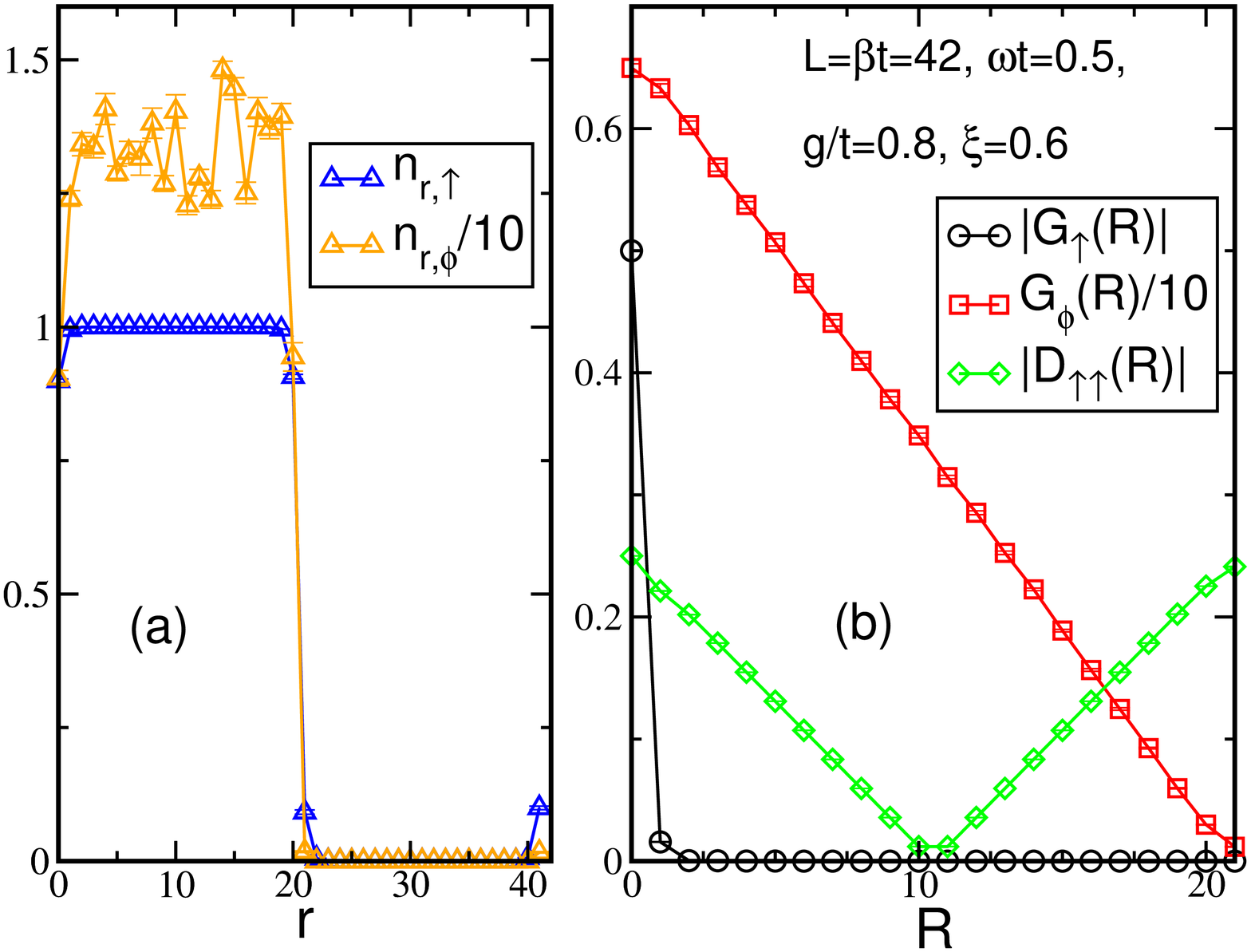}
\caption{(Color online) (a) For $\xi$ large enough, the fermions and
  phonons collapse, forming density plateaus in half the system. (b) In that case, the fermionic Green function,
  $G_\uparrow(R)$, decays exponentially to zero while the
  density-density correlations and phonon Green functions show
  characteristic long range modulations,  due to the plateaus
in the density distributions.
\label{fig:phasesep}}
\end{figure}

Finally, when the range of the coupling is large enough, ($\xi=0.6,
g/t=0.8)$, the Peierls phase is destabilized and the system collapses,
forming a plateau of fermions and phonons surrounded by empty space
(Fig.~\ref{fig:phasesep}~(a)). This happens as the long range
coupling energy overcomes the quantum pressure due to virtual hopping.

The fermionic Green function $G_\uparrow(R)$ decays exponentially, as
the system is either empty or in a state where the movements of the
particles are forbidden by Pauli principle (Fig.~\ref{fig:phasesep}~(b)). The phonons remain coherent throughout the plateau.
$G_\phi(R)$ then shows some long range modulations when averaged over
all starting sites. If the plateau is located between
$r=0$ and $L/2$, $G_\phi(R)$ ($R$ positive and smaller than $L/2$) receives non zero 
contributions from $\langle (a_r a^\dagger_{r+R} + {\rm H.c.})/2\rangle$  only if both $r$ and $r+R$
are located in the plateau, that is if $0\le r<L/2-R$. Each non zero contribution is roughly
equal to the density of phonons in the plateau. Then
$G_\phi(R) \propto (L/2-R)/L = 1/2 - R/L$ decreases linearly with $R$ for $R<L/2$.
The same happens for density-density correlation
functions, as exemplified by $D_{\uparrow\uparrow}(R)$
(Fig.~\ref{fig:phasesep}~(b)).

We see that, in all three phases, the phonons retain some form of
phase coherence, whereas the fermions exhibit quasi-long range
coherence only in the metallic phase. In the following we will use the
stiffness $\rho_{s,\sigma}$ and the behavior of \(G_\sigma(R)\) to 
identify the metallic phase.

\begin{figure}[!h]
\includegraphics[width=1 \columnwidth]{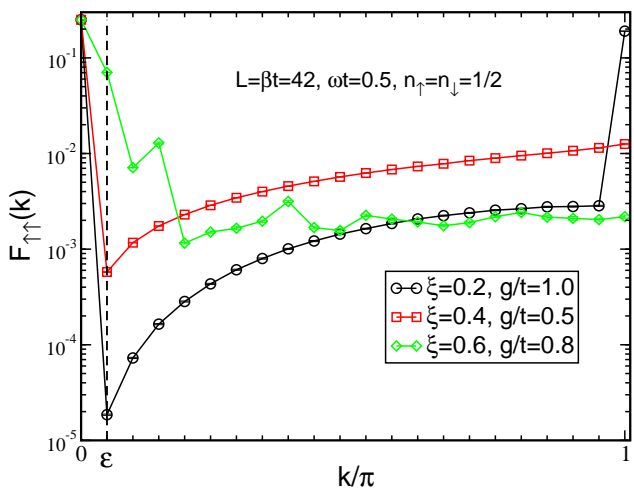}
\caption{(Color online) Comparison of the behaviour of the structure
  factor $F_{\uparrow\uparrow}(k)$ in the three phases. In the
  metallic phase (squares), we only observe a peak at $k=0$
  corresponding to the average density. In the Peierls phase
  (circles), a strong peak appears at $k=\pi$. In the phase separation
  region (diamond), the peak at $k=0$ is enlarged, giving
  significantly larger values for $F_{\uparrow\uparrow}(\epsilon)$
  where $\epsilon = 2\pi/L$.
\label{fig:comparS}}
\end{figure}

To distinguish between the Peierls phase and phase separation, we
consider the behaviour of the structure factors (Fig.~\ref{fig:comparS}).
In all three phases we observe a peak at $k=0$ which simply
corresponds to the average density. As expected, in the Peierls phase,
$F_{\uparrow\uparrow}(k)$ shows a strong peak at $k =\pi$. The
metallic phase shows no particular structure. Finally, the collapsed
phase shows a rather irregular form, which is the effect of the frozen
plateau observed in Fig.~\ref{fig:phasesep}.  However, the long range
modulation induced by the plateau enlarges the $k=0$ peak, which is
not observed in the other phases. As such, the value of
$F_{\uparrow\uparrow}(k)$ for small values of $k$ is much larger for
the collapsed state. Then we can use a large value of
$F_{\uparrow\uparrow}(\epsilon)$, where $\epsilon = 2\pi/L$, as an
indicator that the system has undergone phase separation.

Using these quantities, we build the phase diagram of the system by
doing systematic cuts in the phase space and analyzing the stiffness,
$\rho_{s,\uparrow}$, structure factors, $F_{\uparrow\uparrow}(\pi)$
and $F_{\uparrow\uparrow}(\epsilon)$, as well as the phonon density,
$n_\phi$. We also analyze similar quantities for other types of
particles (down fermions, phonons).

\begin{figure}[!h]
\includegraphics[width=1 \columnwidth]{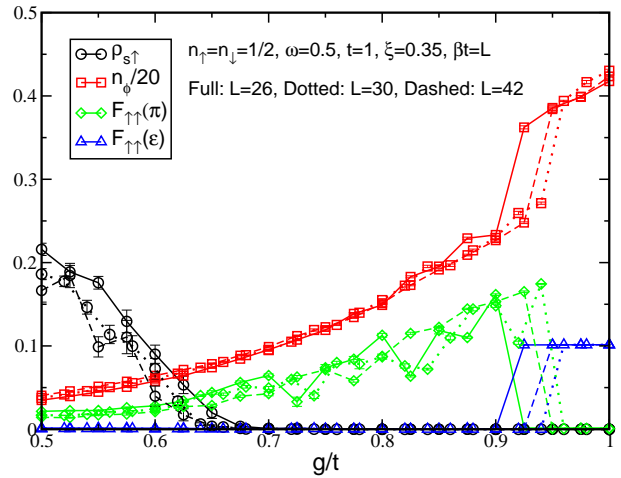}
\caption{(Color online) The stiffness, $\rho_{s\uparrow}$, phonon density,
  $n_\phi$, and structure factor, $F_{\uparrow\uparrow}$ as
  functions of $g$ for $\xi=0.35$. We observe three phases: a metallic
  phase where $\rho_{s,\uparrow}\ne 0$ and $F_{\uparrow\uparrow}(\pi)$
  remains small ($g/t \lesssim 0.65 $); an intermediate Peierls phase
  where $\rho_{s,\uparrow}=0$ and $F_{\uparrow\uparrow}(\pi)$ is
  larger ($0.65\lesssim g/t \lesssim 0.9 $); and phase separation,
  $\rho_{s,\uparrow}= 0$, $F_{\uparrow\uparrow}(\pi) = 0$ and
  $F_{\uparrow\uparrow}(\epsilon) \ne 0$ ($0.9\lesssim g/t$). Upon
  entering the phase separation region, the density of phonons
  $n_\phi$ becomes larger.
\label{fig:cutxifixe}}
\end{figure}

A typical cut, for a fixed value of $\xi = 0.35$, varying $g/t$, and
three sizes $L=26$, 30 and $42$ is shown in
Fig.~\ref{fig:cutxifixe}. We observe successively the three
phases. The metallic phase is characterised by a non zero stiffness
$\rho_{s,\uparrow}$. In the metallic phase $F_{\uparrow\uparrow}(\pi)$
takes a small although non zero value as there are quasi long range
density-density correlations in this phase.  In the Peierls phase,
$\rho_{s,\uparrow}$ is zero and $F_{\uparrow\uparrow}(\pi)$ becomes
larger as there is true long range order for the density-density
correlations. Finally, at large $g$, $F_{\uparrow\uparrow}(\pi)$
becomes suddenly zero while $F_{\uparrow\uparrow}(\epsilon)$
rises. This is accompanied by an abrupt increase of the density of
phonons and signals the occurrence of the phase separation.
Using these signals, we can plot 
the phase diagram shown in Fig.~\ref{fig:phase_diagram_U0}. 
It is quite difficult to locate precisely the
boundaries of the different regions. As can be observed in 
Fig.~\ref{fig:cutxifixe}, the value of $g$ at which the transition from the Peierls
phase to the phase separation regime occurs increases between $L=26$ and $L=30$
and then decreases between $L=30$ and $L=42$, offering no clear systematic scaling behavior.

\begin{figure}[!h]
\includegraphics[width=1 \columnwidth]{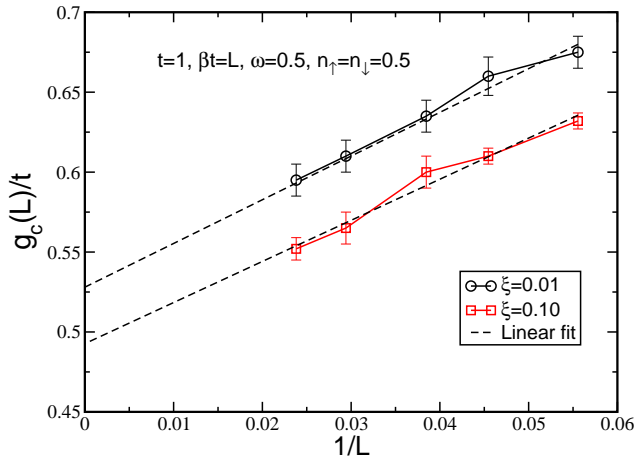}
\caption{(Color online) Finite size scaling extrapolation of the transition point
between the metallic and Peierls phases for low values of $\xi=0.01$ and $\xi=0.1$.
\label{fig:FFSU0} }
\end{figure}

At low $\xi$, the simulations are easier to perform and allow
a finite size scaling analysis. We identified the transition point $g_c$ between the
metal and Peierls phases as the point where $\rho_{s,\uparrow}$ reaches half
its maximum low $g$ value. Plotting $g_c(L)$ as a function of $1/L$ (Fig.~\ref{fig:FFSU0})
we can extrapolate to $L\rightarrow \infty$. For $\xi=0.01\simeq 0$ we find $g_c(\infty)/t = 0.53 \pm 0.03$
which is compatible with previously known results \cite{Fehske08,Hohenadler12}. For $\xi=0.1$ we find $g_c(\infty)/t= 0.49 \pm 0.02$.

We performed  similar analyses with a Langevin algorithm\cite{Batrouni19} and found equivalent results
(see, for example, Fig. \ref{fig:langevin} in appendix \ref{appendix}).

\section{Case with non zero electron-electron interaction}

We now turn to the case where $U \ne 0$. In addition to the three phases
already observed we expect an antiferromagnetic phase to appear in the
system at half-filling.  We will concentrate on the $g=1$ and $g=0.4$
cases as they correspond to the two typical behaviours observed in the
$U=0$ phase diagram (Fig.~\ref{fig:phase_diagram_U0}). In
the first case, we have a transition from a Peierls to phase
separation for $\xi\simeq 0.3$.  In the second, we have a metallic
phase for all the values of $\xi$ we examined, {\it i.e.} up to
$\xi=0.8$.

\subsection{$g=1$, half-filled case}

\begin{figure}[!h]
\includegraphics[width=1 \columnwidth]{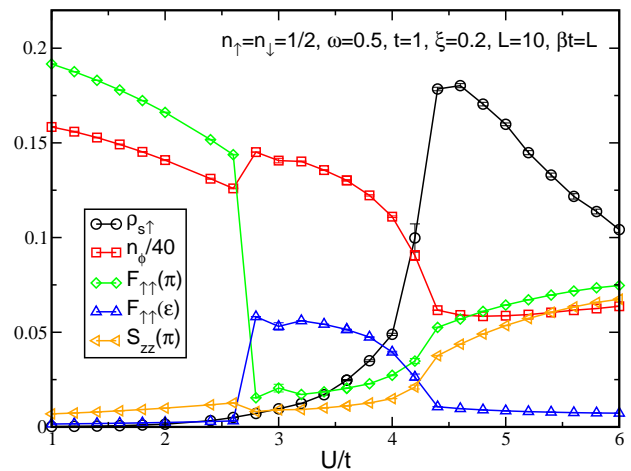}
\caption{(Color online) Cut in the phase diagram for fixed $g=1$ and
  $\xi=0.2$. We observe three different phases. At low $U \lesssim
  2.8$, the system is in a Peierls phase as the only non zero
  structure factor is $F_{\uparrow\uparrow}(\pi)$. A phase separated
  state is found for $2.8 \lesssim U \lesssim 4.2$, marked by the
  larger value of $F_{\uparrow\uparrow}(\epsilon)$. Finally, for
  $U\gtrsim 4.2$ the system is in an antiferromagnetic state
  characterized by $S_{zz}(\pi)$, $F_{\uparrow\uparrow}(\pi)$ and
  $\rho_{s\uparrow}$ being nonzero.
\label{fig:cutxifixeU}}
\end{figure}

Figure \ref{fig:cutxifixeU} shows the evolution of the system for a
fixed $g=1$ and $\xi=0.2$ as $U$ is increased. Starting from a Peierls
phase at $U=0$ (Fig.~\ref{fig:phase_diagram_U0}), we go through a
phase separated state and, finally, an antiferromagnetic Mott phase at
large $U$.  The presence of the AF phase is demonstrated by the fact
that $S_{zz}(\pi)$ is non zero. We also observe that
$F_{\uparrow\uparrow}(\pi)$ and the stiffness $\rho_{s,\uparrow}$ are
non zero in the AF phase, as expected. The change in behaviour of
$n_\phi$ also marks the transitions between the different states, as
$n_\phi$ is generally larger in the collapsed state.

\begin{figure}[!h]
\includegraphics[width=1 \columnwidth]{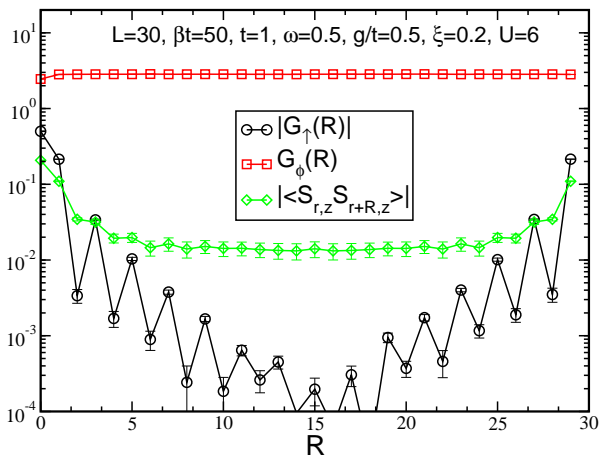}
\caption{(Color online) Behaviour of correlations as functions of
  distance $R$ in the antiferromagnetic Mott phase. $G_\phi(R)$ shows
  the long range phase coherence of the phonons.  $G_\uparrow(R)$
  shows the localisation of the individual fermions in the Mott phase.
  The spin-spin correlations $\langle S_{r,z}S_{r+R,z}\rangle$ decay
  slowly with a behavior that is compatible with long range or quasi
  long range orders.}
\label{fig:spingreen}
\end{figure}

We analyse the AF phase by studying the correlation functions
(Fig.~\ref{fig:spingreen}).  As in the other phases, the phonons
develop a long range phase coherence shown by the behaviour of
$G_\phi(R)$. The fermionic Green function $G_\uparrow(R)$ decays
exponentially with $R$, as expected in a Mott like phase, while the
spin-spin correlations $\langle S_{r,z}S_{r+R,z}\rangle$ reach a
constant value. In one dimension one would rather expect a quasi long
range order with an algebraic decay of $\langle
S_{r,z}S_{r+R,z}\rangle$, because of the continuous symmetry of the spin degrees
of freedom, but it is not visible here, due to
the limited size of the system. A finite size analysis is needed to determine
the exact nature of the spin correlations.

\begin{figure}[!h]
\includegraphics[width=1 \columnwidth]{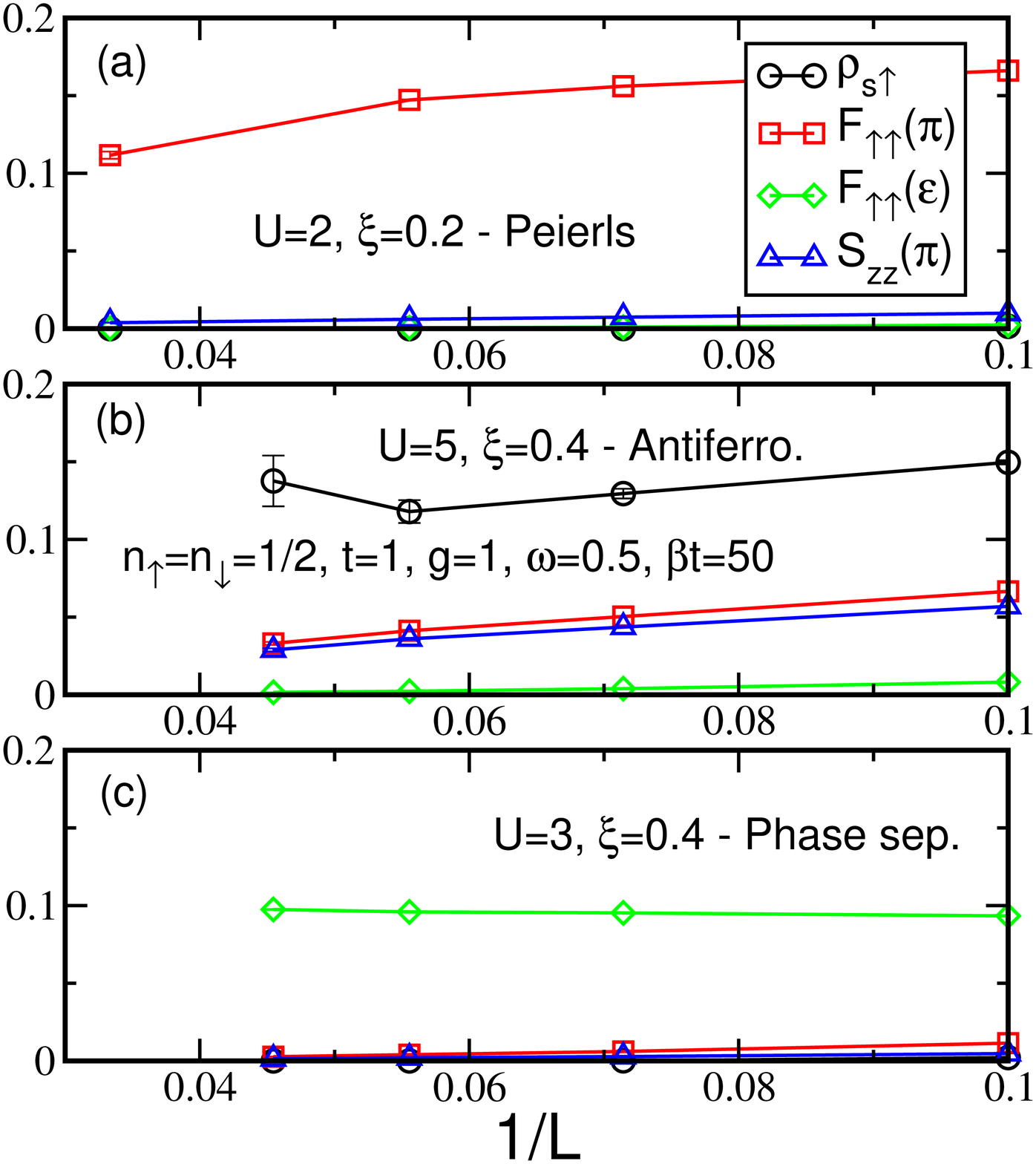}
\caption{(Color online) Finite size behavior of the structure factors
  and stiffness in three different regions: Peierls (a),
  antiferromagnetic (b) and phase separation (c). \label{fig:ffsU}
  While in the Peierls and phase separated state cases
  $F_{\uparrow\uparrow}(\pi)$ and $F_{\uparrow\uparrow}(\epsilon)$,
  respectively, extrapolates to nonzero values when
  $L\rightarrow\infty$, $S_{zz}(\pi)$ and $F_{\uparrow\uparrow}(\pi)$
  extrapolate to a value compatible with zero in the AF
  case.}
\end{figure}

To confirm the presence of three different phases, we perform a finite
size analysis (Fig.~\ref{fig:ffsU}). In the Peierls phase
(Fig. \ref{fig:ffsU}(a)), $F_{\uparrow\uparrow}(\pi)$ extrapolates to
a nonzero value in the large $L$ limit, which signals long range
order, while the other structure factors and the stiffness go to zero.
The same is true for $F_{\uparrow\uparrow}(\epsilon)$ in the phase
separated state (Fig. \ref{fig:ffsU}(c)). In contrast, in the AF
phase (Fig. \ref{fig:ffsU}(b)) the leading structure factors
$S_{zz}(\pi)$ and $F_{\uparrow\uparrow}(\pi)$ decrease with size. A
linear fit gives a value in the $L\rightarrow \infty$ limit that is
compatible with zero.  At the same time, the stiffness
$\rho_{s,\uparrow}$ extrapolates to a non zero value. This is
characteristic of the quasi-long-range AF order that one expects in
one dimension.

\begin{figure}[!h]
\includegraphics[width=1 \columnwidth]{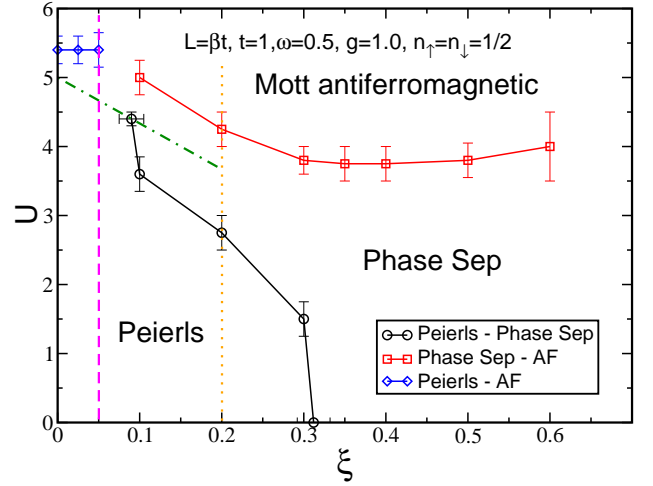}
\caption{(Color online) Phase diagram for $g=1$ and $\omega=0.5$ in
  the half-filled case in the $(\xi,U)$ plane.\label{fig:pdU} We
  observe three phases, a Peierls phase at low $U$ and $\xi$, phase
  separation at low $U$ and large $\xi$ and an AF Mott insulator state
  at large $U$. The dotted line corresponds to the cut shown in Fig.~\ref{fig:cutxifixeU},
the dashed line to Fig.~\ref{fig:cutUsmallxi}, and the dotted dashed line to Fig.~\ref{fig:u5min}.}
\end{figure}

\begin{figure}[!h]
\includegraphics[width=1 \columnwidth]{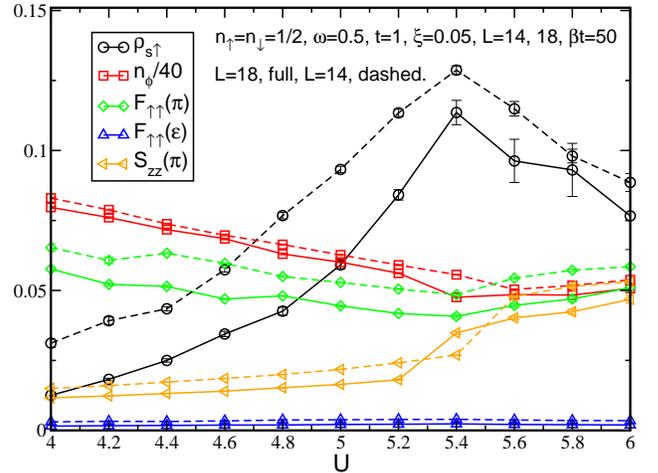}
\caption{(Color online) Cut in the phase diagram at small $\xi=0.05$
  to observe the direct transition from Peierls to
  AF. $F_{\uparrow\uparrow}(\epsilon)$ always remains zero and there
  is no sign of a phase separation. The transition from Peierls to AF
  is difficult to observe but is visible in the small jump in the
  value of $S_{zz}(\pi)$.\label{fig:cutUsmallxi}}
\end{figure}

Using cuts similar to Fig.~\ref{fig:cutxifixeU} for three different
sizes, $L=10, 14$, and $18$, at $\beta t=50$, we draw the phase
diagram for $g=0.1$ in the $(\xi,U)$ plane at half-filling
(Fig.~\ref{fig:pdU}).  We find the three phases presented before.
The phase separated region extends between the Peierls and
AF phase, down to $\xi\simeq 0.1$. To confirm the presence of a direct
Peierls AF transition at low $\xi$ we performed simulations at fixed
small values of $\xi =0, 0.025$ and 0.05 (Fig.~\ref{fig:cutUsmallxi}).
In all these cases, we found that $F_{\uparrow\uparrow}(\epsilon)$
always remains zero, indicating that there is no phase separation and
indeed a direct transition from the Peierls phase to the AF phase. In
the $\xi=0$ limit, our system is the conventional Hubbard-Holstein
model.  We observe a transition from the Peierls to the Mott insulator
for $U\simeq 5.4$ but in this regime, we are limited to small sizes ($L$ up to 18 only). Previous studies
\cite{Hardikar07,Fehske08,Tezuka07} located this transition at a lower
value, slightly above $U=4$. For larger values of $\omega$, there may
be an intermediate metallic phase but this is not the case for
$\omega=t/2$.\cite{Hardikar07,Fehske08,Tezuka07}

\begin{figure}[!h]
\includegraphics[width=1 \columnwidth]{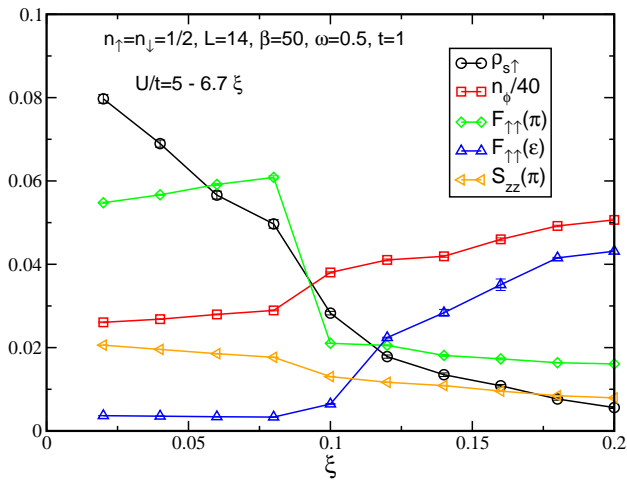}
\caption{(Color online) Cut in the phase diagram along the dotted dashed diagonal line
  in Fig.~\ref{fig:pdU} where $U/t = 5-6.7\, \xi$. We observe a
  transition from Peierls to phase separation.  For $\xi \lesssim
  0.1$, $F_{\uparrow\uparrow}(\epsilon)$ is negligible while
  $F_{\uparrow\uparrow}(\pi)$ is large, which signals the Peierls
  phase. For $\xi \gtrsim 0.1$, $F_{\uparrow\uparrow}(\epsilon)$
  becomes non zero and marks the entry into the phase separated
  region.
\label{fig:u5min}}
\end{figure}

To locate better the left boundary of the phase separation region, we
did some simulations with parameters $U$ and $\xi$ chosen along
diagonal lines in the phase diagram (see the dotted dashed line in Fig.~\ref{fig:pdU}) as shown in Fig.~\ref{fig:u5min}.
We again note an absence of phase separation for $\xi \lesssim 0.1$. As
in the $U=0$ case, the phase separation does not seem to
persist down to $\xi=0$. This is not surprising as the range needs to
be long enough to collapse the system.

\subsection{$g=0.4$, half-filled case}

For a lower value, $g=0.4$, the situation is simpler.  As observed in
the $U=0$ case, for low $g$, the system does not show
phase separation. When electron-electron interactions are increased the system then
simply undergoes a transition from a metallic state to a Mott
antiferromagnet.

\begin{figure}[h]
\includegraphics[width=1 \columnwidth]{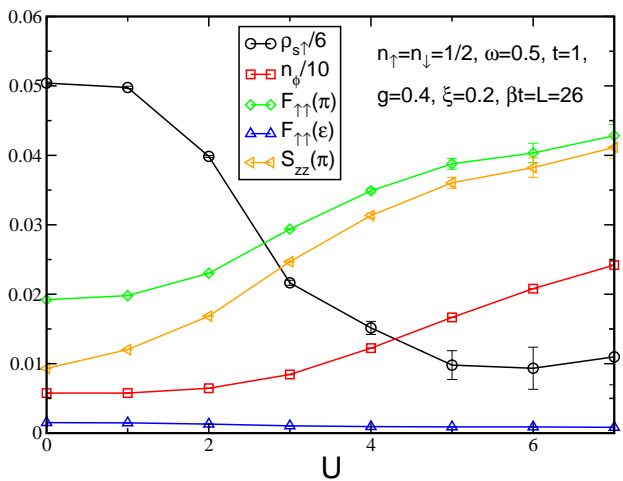}
\caption{(Color online) Structure factors, density of phonons,
  $n_\phi$, and stiffness, $\rho_{s,\uparrow}$, as functions of $U$
  for $g=0.4$. The system goes from a metallic to an AF phase as $U$
  is increased without experiencing phase separation.
\label{fig:cutg04}}
\end{figure}

This is first observed in the evolution of the structure factors in
Fig.~\ref{fig:cutg04}: As $U$ increases, so do $S_{zz}(R)$ and
$F_{\uparrow\uparrow}(\pi)$.  In contrast, the stiffness
$\rho_{s\uparrow}$, while nonzero in both the metallic and AF phases,
drops when one enters the second. 

\begin{figure}[h]
\includegraphics[width=1 \columnwidth]{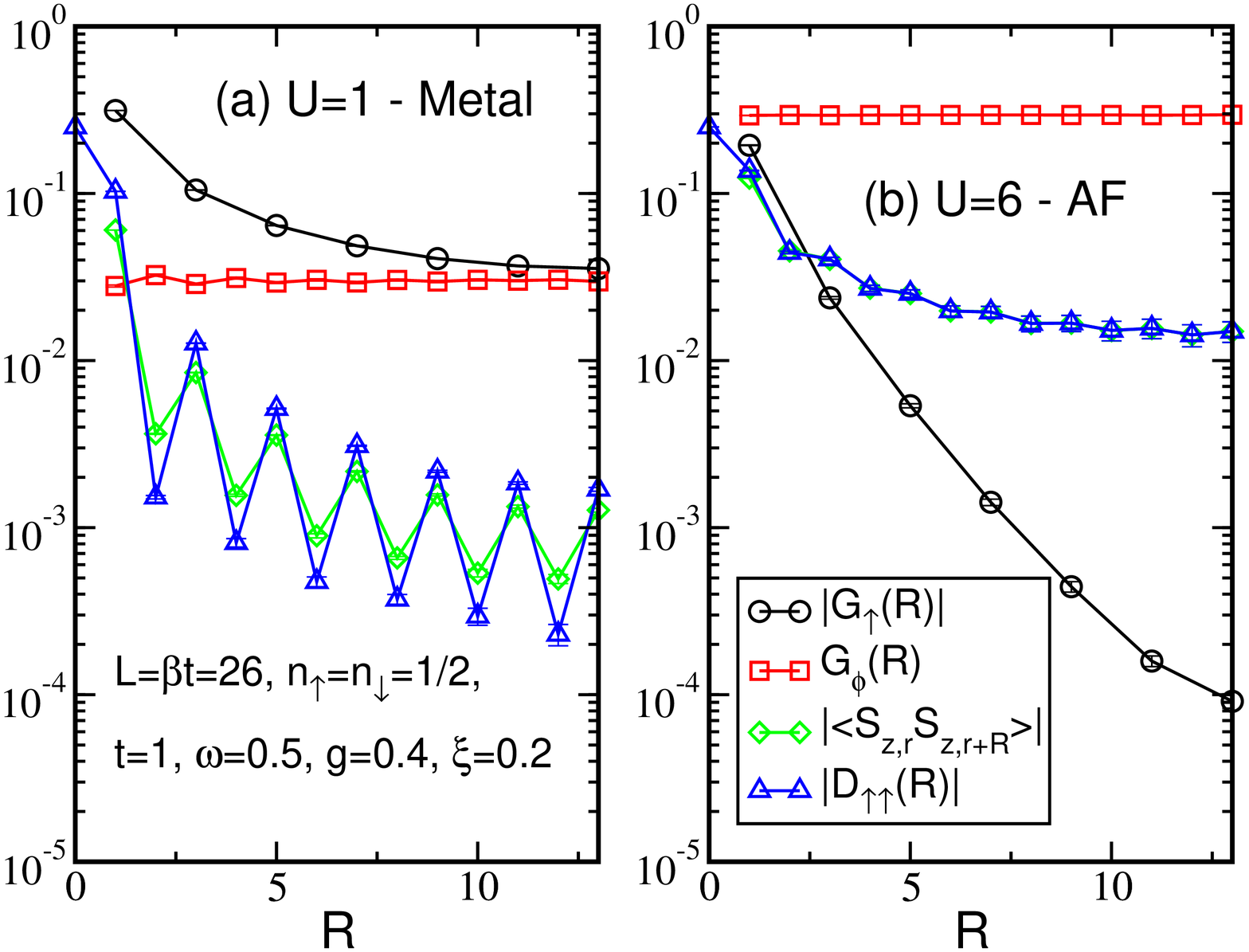}
\caption{(Color online) Correlations as functions of distance for
  $g=0.4$ and $\xi=0.2$ in the metallic phase at low $U=1$ (a) and in
  the antiferromagnetic phase at $U=6$ (b).
\label{fig:corrg04}}
\end{figure}

It is a bit difficult to distinguish the AF phase from the metallic
one in one dimension. Indeed, we do not expect long range magnetic
order for the AF phase and, in one dimension, the metallic phase
should be described by Luttinger physics and also shows some quasi long
range order for the density and spin correlations.

This is indeed what is observed in Fig.~\ref{fig:corrg04} which shows
different correlation functions for weak ($U=1$) and strong ($U=6$)
interactions (Fig.~\ref{fig:corrg04}~(a) and (b), respectively).  For
$U=1$, we observe quasi long range order for the fermion Green
function $G_\uparrow(R)$, the spin-spin correlation $\langle
S_{z,r}S_{z,r+R}\rangle$, and the density-density correlation
$D_{\uparrow\uparrow}(R)$, although the fermion Green function is
clearly the leading correlation in that case. We also observe, as
mentioned before, a true long range order for the phonon phase
coherence $G_\phi(R)$.

On the contrary, for $U=6$ (Fig.~\ref{fig:corrg04}(b)), we observe
that the spin-spin and density-density correlations remain quasi long
ranged while the fermionic Green function decays exponentially, which
is the sign that we are in a Mott insulating phase. This time, the
leading effects are clearly spin correlations. The phonons still show
long range phase order.

\begin{figure}[h]
\includegraphics[width=1 \columnwidth]{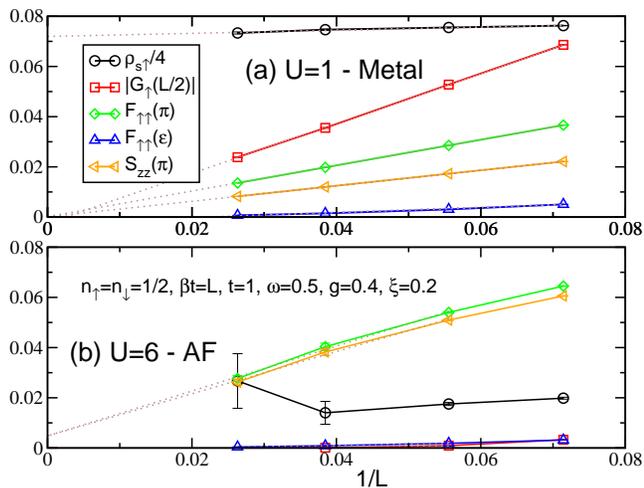}
\caption{(Color online) Scaling of different quantities with size $L$
  for $g=0.4$ and $\xi=0.2$ in the metallic phase at low $U=1$ (a) and
  in the antiferromagnetic phase at $U=6$ (b).  In both, all
  quantities extrapolate to zero except for the stiffness
  $\rho_{s\uparrow}$. In the metal (a), the one body Green
  function $G_\uparrow(R)$ is the leading correlation, whereas in the
  AF (b), $S_{zz}(\pi)$ and $F_{\uparrow\uparrow}(\pi)$ are the
  leading correlations.  The stiffness is difficult to measure in this
  second case.
\label{fig:scaling04}}
\end{figure}

Looking at the scaling of these quantities as a function of size $L$
in the metal $U=1$ phase (Fig.~\ref{fig:scaling04}(a)) and AF phase
(Fig.~\ref{fig:scaling04}(b)) we observe that all quantities scale to
zero, except for the stiffness. This was expected for a one
dimensional system, as all correlation functions show at most quasi
long range order. In the metallic phase, we observe a sizeable one
particle Green function at long distances $G_\uparrow(L/2)$ as well as
noticeable spin-spin $S_{zz}(\pi)$ and density correlations
$F_{\uparrow\uparrow}(\pi)$.  In the AF phase, $S_{zz}(\pi)$ and
$F_{\uparrow\uparrow}(\pi)$ become the leading correlations while
$G_\uparrow(L/2)$ is exponentially suppressed in that case.

\begin{figure}[h]
\includegraphics[angle=-90,width=1 \columnwidth]{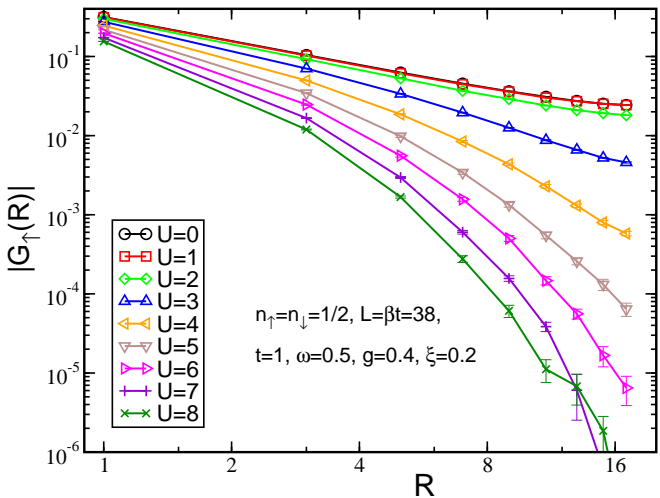}
\caption{(Color online) Behavior of the one particle Green function
  $G_\uparrow(R)$ for $g=0.4$ and $\xi=0.2$ and different values of
  $U$. The behavior changes from an algebraic decay for $U\lesssim 4$
  to an exponential one for $U \gtrsim 4 $.\label{fig:green_g04} }
\end{figure}
As we do not have true long range order in these two phases, the only
behavior that allows their identification is that of the
correlation functions, especially $G_\uparrow(R)$ that changes from an
algebraic to an exponential decay
(Fig.~\ref{fig:green_g04}). Examining simulations on a $L=38$ system,
we find a transition around $U\simeq 4$ for $g=0.4$ and $\xi=0.2$.
This, unsurprisingly, corresponds to the point where the stiffness
$\rho_{s\uparrow}$ drops in Fig.~\ref{fig:cutg04}.

Finally $F_{\uparrow\uparrow}(\epsilon)$ decays rapidly to zero in
both cases, which shows that there is no tendency towards phase
separation for $g=0.4$.

\section{Summary}

We studied a one-dimensional Hubbard-Holstein model with long range
coupling between fermions and phonon, and on-site interaction
between fermions.  The results presented here are limited to the case
of phonon frequency $\omega=t/2$. The physics of
the Hubbard-Holstein with on-site phonon coupling for larger values of
$\omega$ has been studied in\cite{Hardikar07,Fehske08,Tezuka07}.

For $U=0$, the Holstein model, we observed, at half-filling, three
different phases: A metal at low $g$ and, for larger $g$, a transition
from a Peierls CDW phase at small coupling range $\xi$ to a phase
separation region for larger values of $\xi$
(Fig.~\ref{fig:phase_diagram_U0}).  This is reminiscent of the results
found in a previous study \cite{Hohenadler12} although we found the
phase separation region to extend to much smaller values of $\xi$.

Introducing strong enough electron-electron interactions, $U$, drives the half-filled
system towards a Mott antiferromagnetic phase.  For large $g=1$, the
Peierls phase or phase separated region will transform into a Mott for
$U\simeq 5$. However, we observe, as in the non $U=0$ case,
that the phase separation region extends to low $\xi \simeq
0.1$, coming in between the Peierls and Mott phases
(Fig.~\ref{fig:pdU}). A direct Peierls-Mott transition is observed
only for small values of $\xi$.  For $g=0.4$, the metallic phase
present for all studied values of $\xi$ is transformed into an AF Mott phase without an
intermediate phase (Fig.~\ref{fig:cutg04}).

\begin{acknowledgments}
We thank M. Hohenadler and F.F. Assaad for constructive discussion.
The work of B.X. and R.T.S. was supported by DOE grant DE-SC0014671. 
This work was supported by the French government, through the UCAJEDI
Investments in the Future project managed by the National Research
Agency (ANR) with the reference number ANR-15-IDEX-01 and by Beijing
omputational Science Research Center.
\end{acknowledgments}

\appendix

\section{Numerical methods\label{appendix}}

\subsection{Stochastic Green function (SGF) algorithm}

The SGF algorithm, introduced in Refs. \onlinecite{Rousseau08-1,
  Rousseau08}, is a quantum Monte Carlo algorithm which evolved from
the Worm \cite{Proko98} and canonical Worm
algorithms\cite{VanHoucke06}. The main interest of the SGF algorithm
is to allow the measurement of $n$ points equal time Green functions
and the simulation of complex models, especially models that do not
conserve the number of particles.

If the Hamiltonian is written as the sum of two parts $H=V-K$, where
$V$ is diagonal (in a chosen basis) and $K$ nondiagonal, the partition
function can then be expressed as an expansion in powers of $K$
\cite{Proko98}
\begin{equation}
Z= {\rm Tr} \sum_{n=0}^\infty
\int_{0<\tau_1<\tau_2\cdots<\tau_n<\beta} \hspace{-2cm}d\tau_1 \cdots
d\tau_n \ \
e^{-\beta V} K(\tau_n) \cdots K(\tau_1),
\end{equation}
where $K(\tau) = e^{\tau V} K e^{-\tau V}$.  Here we will choose the
occupation number basis. We then have $V = U \sum_r
n_{r,\uparrow}n_{r,\downarrow}+\omega \sum_r n_{r,\phi}$ and $K = t
\sum_{r,\sigma} c_{r,\sigma}^\dagger
(c_{r+1,\sigma}^{\phantom{\dagger}} + {\rm H.c.}) - \sum_{r,R} G(R)
\sqrt{2} X_r n_{r+R}$.

Introducing complete sets of states $|\psi_\tau\rangle$ between
nondiagonal operators, we obtain
\begin{eqnarray}
Z&=& \sum_{n=0}^\infty \sum_{\lbrace|\psi_\tau\rangle
  \rbrace}\int_{0<\tau_1<\tau_2\cdots<\tau_n<\beta} \hspace{-2cm}d\tau_1
\cdots d\tau_n \ \  
\langle \psi_0|e^{-\beta V} K(\tau_n)|\psi_{n-1}\rangle \nonumber \\
&& \times \langle \psi_{n-1}|K(\tau_{n-1}) |\psi_{n-2}\rangle \nonumber
\langle \psi_{n-2}|K(\tau_{n-2}) |\psi_{n-3}\rangle \cdots \nonumber\\
&& \times \langle \psi_{2}|K(\tau_{2}) |\psi_{1}\rangle \nonumber
\langle \psi_{1}|K(\tau_{1}) |\psi_{0}\rangle. \nonumber
\end{eqnarray}
If the product of the matrix elements of the form $\langle
\psi_{k}|K(\tau_{k}) |\psi_{k-1}\rangle$ is positive, it can be used
as a weight with which to sample all the variables
($\lbrace\tau_k\rbrace$, $\lbrace|\psi_k\rangle\rbrace$ and the
expansion order $n$). In practice, we resort to the Jordan-Wigner
mapping of fermions onto hardcore bosons to simulate this
one-dimensional fermionic system and avoid a sign problem for the
weight.

In order to sample $Z$, an extended partition function is introduced
$Z(\tau) = {\rm Tr} e^{-(\beta-\tau) H} {\mathcal G} e^{-\tau H}$ where $\mathcal G$ is
the Green operator defined by
\begin{equation}
{\mathcal G} = \sum_{p,q=0}^\infty w_{pq} \sum_{\lbrace c_k|d_l\rbrace}
\prod_{k=1}^p \frac{b^\dagger_{c_k}}{\sqrt{n_{c_k}+1}}
\prod_{l=1}^q \frac{b_{d_l}}{\sqrt{n_{d_l}+1}}.
\end{equation}
Here the $b^\dagger_c$ operator creates a particle in state $c$.
State $c$ is specified by the type of particle that is created (in our
case, two kinds of hardcore bosons, representing spin up and spin down
fermions, or phonons) and by the site on which it is created. The
$b_d$ operator destroys a particle in state $d$, in the same way. In
the Green operator, the $c$ and $d$ states should be different, so
that there are no diagonal contributions in the Green operator, except
for $q=p=0$ which gives the identity operator.  As the terms in $\mathcal G$
are products of creation and destruction operators, $\mathcal G$ is then the
sum of all possible $n$-point Green functions, weighted by the matrix
$w_{pq}$. The Green functions that have large weights $w_{pq}$ will
appear more often is the sampling of $Z(\tau)$.

$Z(\tau)$ is expressed in the same way as $Z$, introducing an
additional set of complete states,
\begin{eqnarray}
Z(\tau)&=& \sum_{n=0}^\infty \sum_{\lbrace|\psi_\tau\rangle
  \rbrace}\int_{0<\tau_1<\tau_2\cdots<\tau_n<\beta} \hspace{-2cm}d\tau_1
\cdots d\tau_n \ \
\langle \psi_0|e^{-\beta V} K(\tau_n)|\psi_{n-1}\rangle  \nonumber
\\  
&&\times \cdots \times \langle \psi_{L+1}|K(\tau_L) |\psi_{L}\rangle
\times \langle \psi_{L}|{\mathcal G}(\tau)|\psi_{R}\rangle \nonumber\\&& \times
\langle \psi_{R}|K(\tau_{R}) |\psi_{R-1}\rangle\times \cdots \times
\langle \psi_{1}|K(\tau_{1}) |\psi_{0}\rangle,
\end{eqnarray}
where we used labels $L$ and $R$ to denote the states appearing on the
left and right of $\mathcal G$.

Whenever $|\psi_L \rangle = |\psi_R\rangle$ during the sampling, the
contribution of the Green operator is the simple constant
$w_{00}$. The configuration that is then obtained by sampling
$Z(\tau)$ also contributes to the original partition function $Z$.
When $|\psi_L \rangle \ne |\psi_R\rangle$, only one of the terms
present in $\mathcal G$ gives a non zero contribution to $\langle
\psi_{L}|{\mathcal G}|\psi_{R}\rangle$. In that case, the configuration obtained
contributes to the sampling of one peculiar Green function.
\begin{figure}[h]
\centerline{\includegraphics[width=0.8\columnwidth]{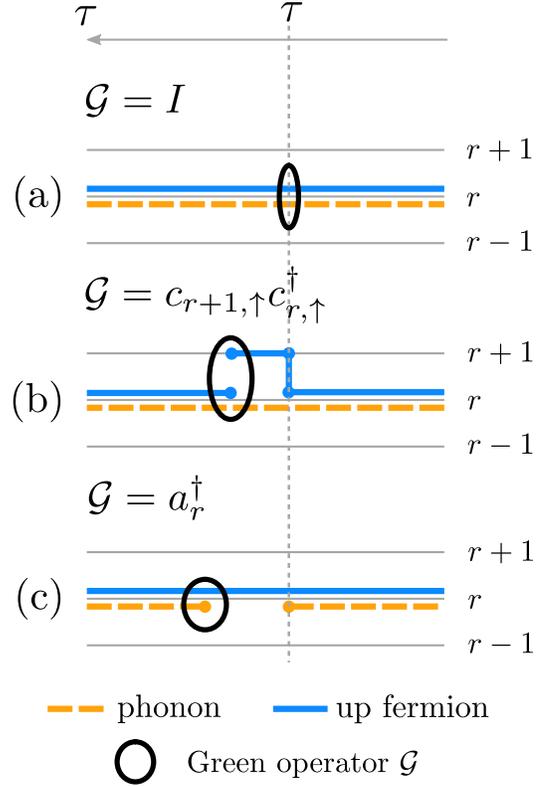}}
\caption{(Color online) Schematic representation of the movement of
  the Green operator ${\mathcal G}$.  Starting from an initial configuration (a)
  where ${\mathcal G}=I$ and where there is one up particle (blue thick line) and
  a phonon (orange thick dashed line) on site $r$, $\mathcal{G}$ is shifted to
  the left ($\tau$ increasing) and a $K$ operator is created.  In (b),
  a jump of the particle has been created, which modified the state on
  the right of $\mathcal G$ and the $\mathcal G$ operator itself. In (c) a destruction
  operator for the phonon has been created and $\mathcal G$ is a phonon
  creation operator.\label{fig:sgf}}
\end{figure}

In practice, the sampling of the extended partition function is made
by using the Green operator. In the simplified update scheme
introduced in \cite{Rousseau08}, two possible ``movements" of $\mathcal G$ are
shown to allow an ergodic sampling of the configurations. First a
shift direction is chosen for ${\mathcal G}(\tau)$ (left if $\tau$ is increased,
right if $\tau$ is decreased). Then moving in this direction, two
different situations can occur: the Green operator can create a $K$
operator at its imaginary time and then be shifted, or the Green
operator can be shifted to the imaginary time of the next $K$ operator
and destroy it.  Creating a $K$ operator requires to choose a new
$|\psi_R\rangle$ state, assuming that a left move is chosen.
Depending on the chosen $|\psi_R\rangle$, the Green operator is
modified accordingly and only one of the terms appearing in $K$ gives
a non zero contribution. The choice between all possible new
$|\psi_R\rangle$ is made with a probability chosen to respect detailed
balance. For example, in our case, the $K$ operator comprises two
kinds of operators: jumps of particles from one site to the next or
creation or destruction of a phonon. The creation of such operators
and the corresponding modifications of the states and Green operator
are illustrated in Fig. \ref{fig:sgf}. The destruction of a $K$
operator modifies in the same way $\mathcal G$ and the states.

The Green operator is moved until it becomes the identity operator, at
which point the measurement of diagonal quantities can be
performed. To sample efficiently $Z(\tau)$, a directed propagation
\cite{Rousseau08} is generally used to avoid $\mathcal G$ going back and forth
in imaginary time. In that case, there is a stronger probability for
the operator to continue its movement in the same direction as in the
previous step.

\subsection{Langevin algorithm}

The Langevin method we used is introduced and benchmarked in
\cite{Batrouni19} where some additional results for $U=0$ are also
presented.

The method initially proceeds in a way that is similar to a
determinant quantum Monte Carlo method \cite{dqmc} (DQMC).  We first
rewrite the phonon diagonal energy of Hamiltonian (\ref{Hamiltonian})
as $\omega n_{r,\phi} = \omega x_r^2/2 + p_r^2/2$ and add a chemical
potential term $-\mu \sum_r n_r$ to the Hamiltonian as the algorithm
works in the grand canonical ensemble.  When $\mu = -[\sum_R
  G(R)]^2/\omega$, the resulting Hamiltonian is particle-hole
symmetric and $\langle n_\uparrow \rangle = \langle n_\downarrow
\rangle = 1/2$.

The partition function is written as a discrete path integral, where
inverse temperature $\beta$ is divided into $L_\tau$ steps of size
$\Delta \tau = \beta/L_\tau$ and complete sets of states $\lbrace
x_{r,\tau}, p_{r,\tau}\rbrace$ are introduced at each imaginary time
step $\tau$.  When $U=0$, the fermionic terms in the Hamiltonian
(\ref{Hamiltonian}) are quadratic and can be traced out, and the
momentum dependence of the phonons can be integrated out, leading to
an expression of the partition function that depends only on the
phonon field $x_{r,\tau}$ \cite{Scalettar89}
\begin{eqnarray}
Z &=& \int \mathcal{D}x_{r,\tau} \exp(-S_{\rm Bose}(\lbrace
x_{r,\tau}\rbrace))\nonumber
[\det M(\lbrace x_{r,\tau}\rbrace)]^2\\ & =& \int
\mathcal{D}x_{r,\tau} \exp(-S(\lbrace
x_{r,\tau}\rbrace)).
\label{eq:langevin} 
\end{eqnarray}
Detailed expressions for $S_{\rm Bose}$ and matrix $M$ are found in
\cite{Batrouni19,dqmc}.  $M$ is a large sparse matrix of dimension
$LL_\tau$ and the method is free of the sign problem as the
determinant of $M$ is squared.

\begin{figure}[h]
\centerline{\includegraphics[width=1\columnwidth]{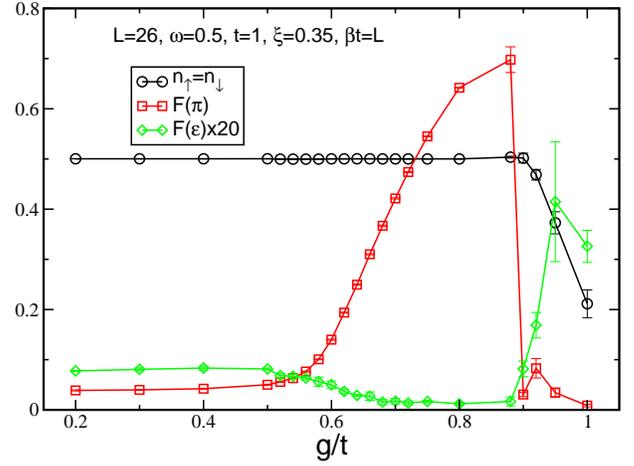}}
\caption{(Color online) Langevin simulations results for a cut in the
  phase diagram using the same parameters as in
  Fig. \ref{fig:cutxifixe}. Chemical potential is set to $\mu =
  -[\sum_R G(R)]^2/\omega$ to impose particle hole symmetry. $F(k)$ is
  the Fourier transform of $\langle n_r n_{r+R}\rangle$.  We observe a
  homogeneous metallic phase for $g/t \lesssim 0.6$ where the density
  structure factors are small. A Peierls phase is observed for
  $0.6\lesssim g/t \lesssim 0.9$, marked by a large value of
  $F(\pi)$. Finally a phase separated region is obtained for $g/t
  \gtrsim 0.9$ as shown by the increase of $F(\epsilon)$ and by the
  average densities $\langle n_\uparrow \rangle = \langle
  n_\downarrow\rangle$ that are no longer 1/2.\label{fig:langevin}}
\end{figure}

The algorithm then proceeds by using a fictitious stochastic dynamics,
governed by the Langevin equation
\begin{equation}
\frac{d x_{r,\tau}(t)}{dt} = - \frac{\partial S}{\partial
  x_{r,\tau}(t)} + \sqrt{2}\eta_{r,\tau}(t),
\end{equation}
where $\eta_{r,\tau}(t)$ are stochastic variables satisfying
$$ \langle \eta_{r,\tau}(t)\rangle =0, \ \langle
\eta_{r,\tau}(t)\eta_{r',\tau'}(t')\rangle
=\delta_{r,r'}\delta_{\tau,\tau'} \delta(t-t').
$$
The Langevin dynamics assures that, when $t\rightarrow \infty$,
variables are distributed according to $P=\exp(-S)$.

Two main technical difficulties need to be overcome in order to
integrate the Langevin equations efficiently.  First, calculating
$\partial S/\partial x_{r,\tau}(t)$ involves \cite{Batrouni19} a trace
over an expression containing the inverse of matrix $M$. This would be
extremely taxing in terms of simulation time, as inverting a matrix
scales as the cube of its dimension.  This trace is then calculated
using a stochastic estimator which allows to replace the matrix
inversion problem with a much simpler solution of a linear
system. This solution is obtained by a conjudate gradient method that
scales linearly with the dimension of $M$, $LL_\tau$. This is a big
advantage of this method as other techniques, such as the conventional
DQMC algorithm, scale as $L^3 L_\tau$.  The second difficulty comes
from the autocorrelation times of the Langevin dynamics, which are
generally very long.  This is solved by the so-called Fourier
acceleration of the Langevin dynamics\cite{Batrouni19}.

\subsection{Langevin simulations results}

Using this algorithm, we confirmed the results obtained with the SGF
algorithm at $U=0$, especially the fact that a phase separation is
present for small values of $\xi$. Fig. \ref{fig:langevin} shows a cut
in the phase diagram for the same parameters as in
Fig. \ref{fig:cutxifixe}.  As with SGF simulations, we observe a
transition from the metal to the Peierls phase for $g/t \simeq 0.6$
and from the Peierls phase to a phase separated behavior for
$g/t\simeq 0.9$. The Peierls phase is here signalled by the large
value of $F(\pi)$ in the intermediate region, where $F(k)$ is the
Fourier transform of $\langle n_r n_{r+R}\rangle$.  The phase
separation is marked by the increased value of $F(\epsilon)$ and,
because the simulations are performed in the grand canonical ensemble,
by the fact that the densities $\langle n_\uparrow \rangle = \langle
n_\downarrow\rangle$ departs from their expected value of 1/2. Indeed,
the density of particles in the system becomes arbitrary in the phase
separation region despite the fact that the chemical potential has
been chosen to ensure particle hole symmetry.

\end{document}